\documentclass[multphys,vecphys]{svmult}

\usepackage{makeidx}     
\usepackage{graphicx}    
\usepackage{multicol}    

\makeindex             

\begin{document}

\title*{The Merger QSO 3C 48 and its Host Galaxy in the Near Infrared
}

\author{J. Zuther\inst{1}, A. Eckart\inst{1}, J. Scharw\"achter\inst{1}, M. Krips\inst{1}, S. Pfalzner\inst{1}, \and C. Straubmeier\inst{1}}

\institute{1. Physikalisches Institut, Universit\"at zu K\"oln,
Z\"ulpicher Str. 77, 50937 K\"oln
\texttt{zuther@ph1.uni-koeln.de}}
\authorrunning{J. Zuther at al.}

\maketitle

In this contribution we present new near-infrared (NIR) data on the quasar 3C 48 and its host galaxy, obtained with ISAAC at the Very Large Telescope (European Southern Observatory, Chile). The NIR images and spectra reveal a reddening of several magnitudes caused by extinction due to molecular material and dust within the host galaxy. For the first time we clearly identify the highly reddened potential second nucleus 3C 48A about $1''$ northeast of the quasar position in the NIR. Its reddening can be accounted for by warm dust, heated by star formation or an interaction of the 3C 48 radio jet with the interstellar medium, or both. The NIR colors and the CO(6-3) absorption feature both give a stellar contribution of about 30\% to the QSO-dominated light. These results will contribute to the question of how the nuclear activity and the apparent merger process are influencing the host galaxy properties and they will improve existing models.

\section{Introduction}
\label{sec:1}
A key goal in understanding the connection between nuclear activity (accretion onto a super massive black hole) and host galaxy structure (e.g. morphology, stellar populations, etc.) is to find out about the physical conditions of quasars and the galaxy hosts they reside in.

Because of its large angular size and its brightness at the redshift of $z\approx 0.37$ \cite{boroson84}, 3C 48 is an ideal target for high-resolution imaging and spectroscopy with large telescopes. The near-infrared (NIR) is sensitive for the mass dominating stellar populations and at the same time it is less affected by extinction than the optical.
The radio-loud galaxy shows an excess far-infrared (FIR) emission, $L_\mathrm{FIR} =5\times10^{12} L_\odot$ \cite{neug85}, and a one-sided radio jet extending about $0.5''$ to the north, fanning out to the east to about $1''$ \cite{wilk91}. 
The quasar is rich in molecular gas, $M_{\mathrm{H}_2}\sim  2.7\times 10^{10} M_\odot$, as CO(1-0) observations show \cite{scoville93, wink97}.
There is also morphological evidence for a recent merger event. Such are a possible double nucleus and a tidal tail extending several arcseconds to the northwest \cite{stockton91, chatzi99, canalizo2000}. Yet it is still unclear, whether the brightness peak about $1''$ northeast (NE) of the QSO is really the nucleus of the companion galaxy, or a region of interaction of the radio jet with the interstellar medium of the host galaxy inducing star formation \cite{canalizo2000}.

\section{Near Infrared Properties of 3C 48}
\label{sec:2}
$J$, $H$, and $Ks$ band imaging as well as $H$ and $K$ band spectroscopy were carried out using the NIR camera-spectrometer ISAAC at the Very Large Telescope (VLT, ESO, Chile) \cite{zuther03}.

\subsection{$JHKs$ Photometry}
\label{sec:2.1}
\begin{figure}[ht!]
\centering
\includegraphics[width=5.7cm]{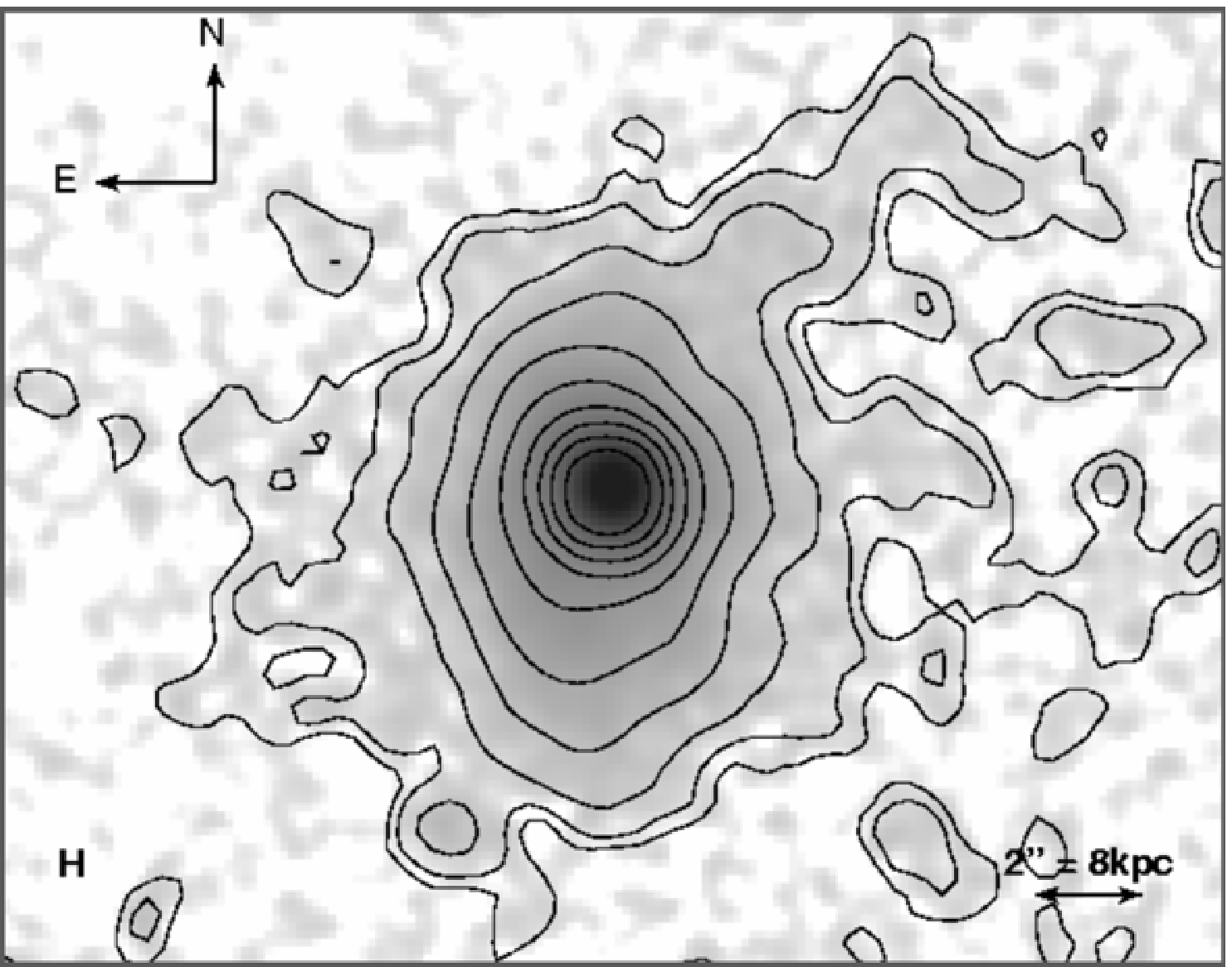}
\includegraphics[width=4.4cm]{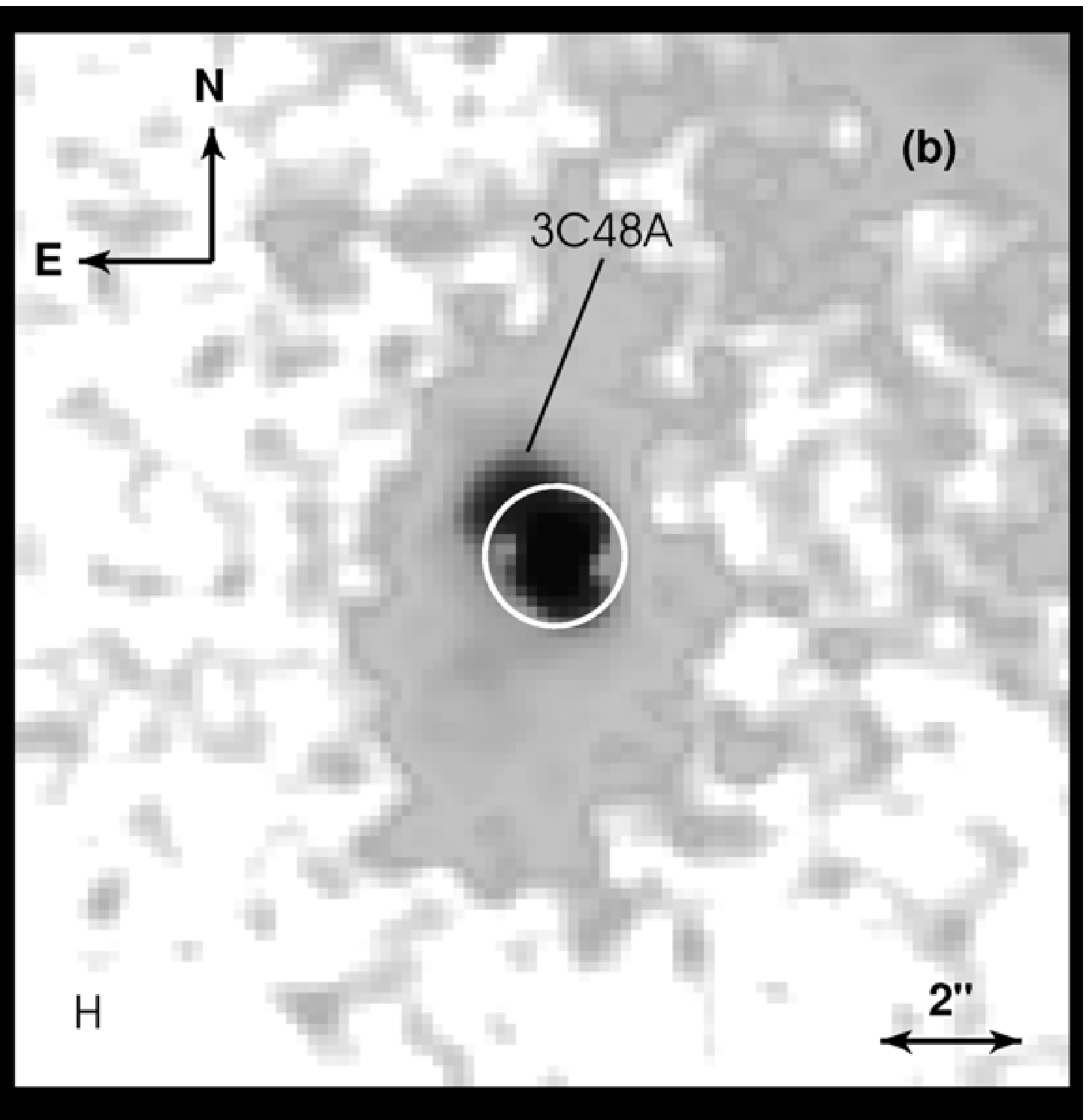}
\caption{ISAAC $H$ band image of 3C 48. \emph{Left:} Original image, still being dominated by the QSO. \emph{Right:} Core-subtracted image, emphasizing the excess brightness of 3C 48A. Both taken from \cite{zuther03}.}
\label{fig:1}       
\end{figure}

\begin{figure}[ht!]
\centering
\includegraphics[width=5cm]{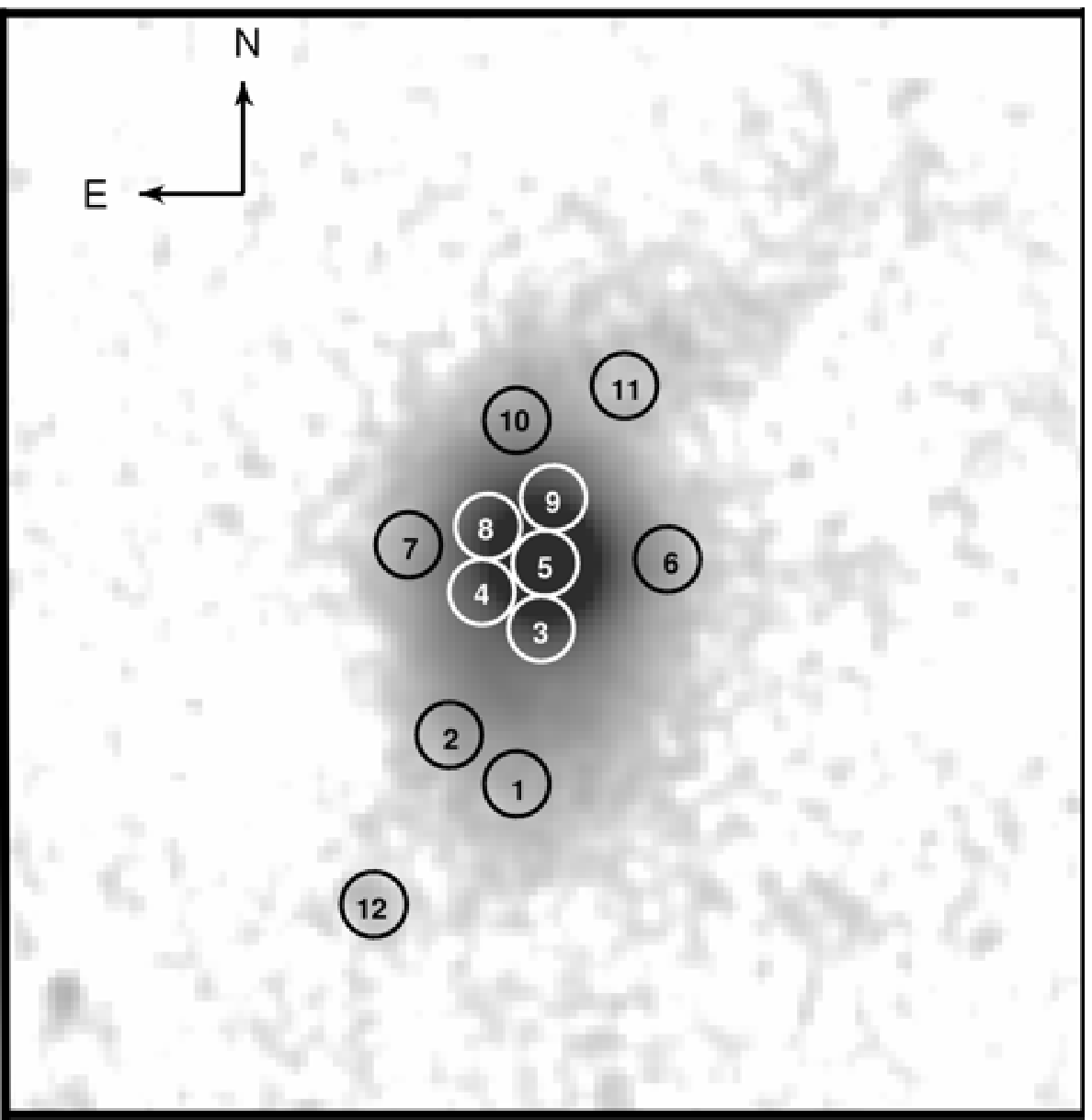}
\includegraphics[width=9.5cm]{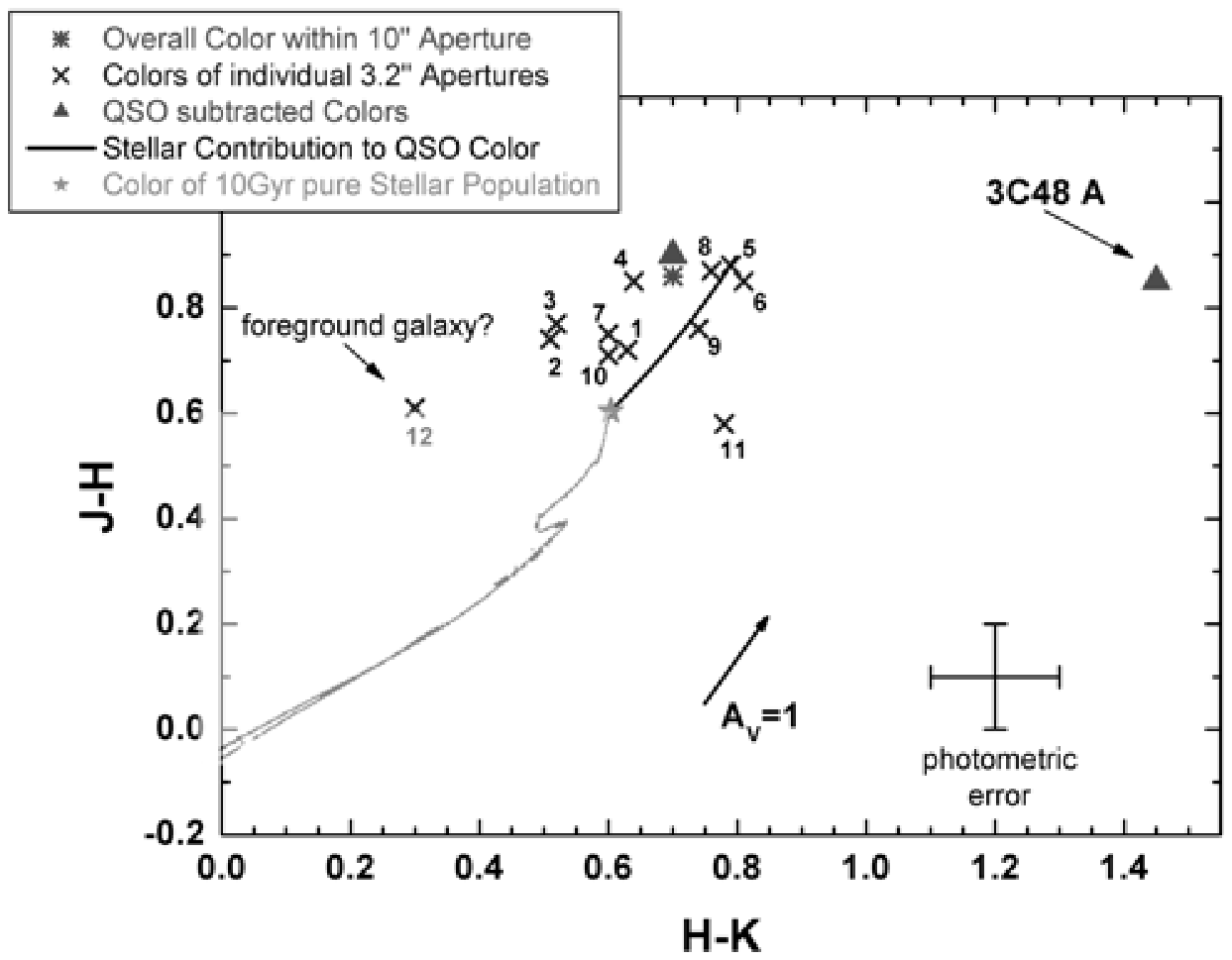}
\caption{\emph{Top panel:} $H$ band image of 3C 48 showing the loci of the $3.2''$ apertures for measuring colors. \emph{Bottom panel:} NIR two color diagram. The symbols are explained in the figure. The light solid line represents an evolutionary model for colors of a single, passively evolving starburst \cite{hutch97}. Evolution proceeds from the lower left corner at 0~Gyr to the upper right corner at 20~Gyr (10 Gyr are marked with a $\star$). The dark solid line represents the stellar contribution to the QSO light when increasing the aperture. The reddening vector for 1 mag of visual extinction is also shown. Both panels taken from \cite{zuther03}.}
\label{fig:2}       
\end{figure}

The images at 1.2, 1.65, and 2.2~$\mu$m ($J$, $H$, and $Ks$) show a
morphology quite similar to the optical one \cite{stockton91}. A tidal tail like structure extends to the northwest of the quasar (Fig. \ref{fig:1}).
Another feature in the southeast of the quasar was previously thought to be a counter tidal tail \cite{canalizo2000}. Our two-color analysis of the host galaxy showed that this feature is most probable not associated with 3C 48. Instead, with colors typical for low-redshift galaxies, it seems to be a foreground galaxy (Fig. \ref{fig:2}, aperture no. 12).

The two-color diagram in Fig. \ref{fig:2} shows that all
colors of the host galaxy are reddened with respect to an old
(10~Gyr) single starburst population (star). The nucleus
itself is also heavily reddened (aperture no. 5; obscuration by
circum nuclear material). The grey curve presents the color evolution
of a single starburst population from zero (lower left
end) to 10~Gyr. Young starbursts therefore have quite blue
colors. It has been shown that there is a lot of star formation
going on in the host \cite{canalizo2000}. Starbursts have quite blue NIR colors (cf. evolutionary curve in Fig. \ref{fig:2}). Therefore a huge extinction seems
to be at work (all data points are redder than the 10 Gyr color
point).

In order to study the nuclear region, especially 3C 48A, the nuclear contribution has to be removed. This was done by subtracting a nearby star, scaled to the peak flux of the QSO.
The result is presented in the right panel of Fig. \ref{fig:1}, showing an
excess brightness about $1''$ northeast of the QSO. The colors
of this region are strongly reddened due to the presence of warm/hot dust (only in the $H-K$ direction).
Furthermore there appears to be an alignment of the radio jet with 3C 48A in the NIR. However, multi-particle simulations by \cite{scharw03} show, that the A component could still be the nucleus of the companion galaxy.
\subsection{$HK$ Spectroscopy}
\label{sec:2.2}
Emission line spectroscopy allows to probe the interstellar
medium (ISM) in the central region of the galaxy, e.g. via the
recombination lines Pa$\beta$ and Pa$\gamma$. In our case we find a line ratio
of about $1.6(\pm 30\%)$. This is a value typical for the ISM of active
galaxies being excitated by the central-engine continuum emission ($T \sim 10$~ K and $n \sim 10$~cm$^{-3}$).

Absorption lines can be used to trace the stellar content. The depth of the CO(6-3) absorption line is a measure for a population of cool old stars \cite{origlia93}. 
In our case the depth of the line measured only as an upper limit corresponds to a stellar contribution of about 30\%. This is consistent with the results from photometry. For details see \cite{zuther03}.

\section{3C 48 revisited}
\label{sec:3}
Our NIR imaging and spectroscopy result in a first detection of 3C 48A in $J$, $H$, and $Ks$. The host galaxy, but especially 3C 48A, are strongly reddened. Both scenarios, a second nucleus or an interaction of the ISM with the radio jet, are still possible to explain the nature of 3C 48A. The presence of significant extinction is an important ingredient for future studies of the stellar populations of the host galaxy. Multi-particle simulations by \cite{scharw03} can explain the missing counter tidal tail as the result of a projection effect, consistent with the optical \cite{canalizo2000} and our NIR data. The merger morphology, its high FIR luminosity, and its large content of molecular gas put 3C 48 in an evolutionary scheme envisioned by \cite{sander88}. There it fits nicely as a transition objects between an ultra-luminous infrared galaxy (ULIRG) phase and a pure QSO phase, being still dominated by its FIR excess but also displaying QSO features like its luminosity and rather flat spectral energy distribution.

%
%

%

\begin{thebibliography}{99.}
\bibitem{boroson84}T. A. Boroson \& J. B. Oke: ApJ \textbf{281}, 535 (1984)
\bibitem{neug85}G. Neugebauer and B.~T. Soifer: ApJL \textbf{295}, L27 (1985)
\bibitem{wilk91}P. N. Wilkinson, A. K. Tzioumis, J. M. Benson, et al.: Nature \textbf{352}, 313 (1991)
\bibitem{scoville93}N. Z. Scoville, S. Padin, D. B. Sanders, B. T. Soifer, \& M. S. Yun:  ApJ \textbf{415}, L75 (1993) 
\bibitem{wink97}J. E. Wink, S. Guilloteau, \& T. L. Wilson: A\&A \textbf{322}, 427 (1997)
\bibitem{stockton91}A. Stockton \& S. E. Ridgway: AJ \textbf{102}, 488 (1991)
\bibitem{chatzi99}E. T. Chatzichristou, C. Vanderriest, \& W. Jaffe: A\&A \textbf{343}, 407 (1999)
\bibitem{canalizo2000}G. Canalizo and A. Stockton: ApJ \textbf{528}, 201 (2000)
\bibitem{zuther03} J. Zuther et al.: submitted to A\&A (2003)
\bibitem{origlia93}L. Origlia, A. F. M. Moorwood,\& E. Oliva:  A\&A \textbf{280}, 536 (1993)
\bibitem{sander88}D. B. Sanders, B. T. Soifer, J. H. Elias, et al.: ApJ \textbf{325}, 74 (1988)
\bibitem{hutch97}J. B. Hutchings \& S. G. Neff: AJ, \textbf{113}, 550 (1997)
\bibitem{scharw03}J. Scharw\"achter et al.: astro-ph/0310740

\end{thebibliography}
%



\printindex
\end{document}